\documentclass[12pt,epsf]{article}
\usepackage{epsfig}
\usepackage{amssymb}  
\usepackage{amsthm}
\usepackage{amsmath}
\let\eqref\cref
\newcommand{\onefigure}[2]{\begin{figure}[htbp]
\begin{center}\leavevmode\epsfbox{#1.eps}\end{center}\caption{#2\label{#1}}
\end{figure}}

\renewcommand{\thanks}[1]{\footnote{#1}} 

\newcommand{\be}{\begin{equation}}
\newcommand{\ee}{\end{equation}}
\newcommand{\bea}{\begin{eqnarray}}
\newcommand{\eea}{\end{eqnarray}}
\newcommand{\half}{{1\over 2}}

\def\stacksymbols#1#2#3#4{\def\theguybelow{#2}
\def\verticalposition{\lower#3pt}
\def\spacingwithinsymbol{\baselineskip0pt\lineskip#4pt}
\mathrel{\mathpalette\intermediary#1}}
\def\intermediary#1#2{\verticalposition\vbox{\spacingwithinsymbol
\everycr={}\tabskip0pt
\halign{$\mathsurround0pt#1\hfil##\hfil$\crcr#2\crcr
\theguybelow\crcr}}}

\newcommand{\goesto}[2]{\stacksymbols{\Longrightarrow}{{}_{#1 \rightarrow #2}}{4}{0.1} }

\begin{document}

\pagestyle{empty}

\bigskip\bigskip
\begin{center}
{\bf \large Unitary Kinematic Mixing of Electro-Weak Bosons}
\end{center}

\begin{center}
James Lindesay\footnote{e-mail address, jlindesay@howard.edu} \\
Computational Physics Laboratory \\
Howard University,
Washington, D.C. 20059 
\end{center}
\bigskip

\begin{abstract} 
In the standard model, electro-weak bosons are developed as gauge-fields initially satisfying
an SU(2) $\times$ U(1) local symmetry mediating interactions with  a multi-component field.
This symmetry gets broken when the component of the field associated with the (scalar) Higgs boson
acquires a non-vanishing expectation value, generating masses for weak vector bosons (W$^+$,W$^-$,Z) as well as a massless
photon (A) which couples to electric charge.  Beyond an expression relating the masses of the W and Z bosons,
there are few predictions concerning further relationships involving the Higgs mass or mixing angle.
In this paper, kinematic relationships between unitarily mixed momentum-space spinors 
(supplemental to standard electro-weak modeling) will be developed. 
These relationships expressed in terms of kinematic invariants are shown to be
consistent with current measured results.

\end{abstract}

\bigskip 

\setcounter{equation}{0}
\section{Introduction}
\indent \indent

Quantum fields representing fundamental particles are generally required to be
eigenstates of the operator $ \hat{P}_\mu \eta^{\mu \nu}  \hat{P}_\nu$, with eigenvalues given
by $-(m c)^2$.  Field equations utilizing operator forms that are linear in the four-momentum have
analytic behaviors that are directly manipulable for developing straightforward relationships
using standard scattering theory, especially with regards to cluster decomposition
properties.  In a generalization of Dirac's formulation\cite{Dirac}, spinor fields
that require the form $\hat{\Gamma}^\mu \: \hat{P}_\mu$  to
be a Lorentz scalar operation, while satisfying
\be
\mathbf{\Gamma}^\beta \cdot {\hbar \over i} { \partial \over \partial x^\beta} \,
\hat{\mathbf{\psi}}_{\gamma}^{(\Gamma)}
(\vec{x}) = -\gamma m  c  \: \hat{\mathbf{\psi}}_{\gamma}^{(\Gamma)}(\vec{x}) ,
\label{CausalSpinorFieldEqn}
\ee
can be developed.  For these fields,  $m \ge 0$ always,
$\mathbf{\Gamma}^\beta$ are the
finite dimensional matrix representations of the operators $\hat{\Gamma}^\beta$,
and for massive states, $\gamma$ are the eigenvalues of the hermitian operator  $\hat{\Gamma}^0$.
In (\ref{CausalSpinorFieldEqn}), the particle mass  $m$
labels a unitary representation of the (extended) Poincare group of transformations
on the particle state.
The fundamental Dirac representations are fermion fields with $\Gamma=\half$, where
the standard Dirac matrices $\gamma^\beta$ satisfy $\mathbf{\Gamma}^\beta=\half \gamma^\beta$.

Standard model
electro-weak bosons originate from massless (configuration-space) gauge fields, with charged components ($W^\pm$) which
become massive (while maintaining renormalizability) via interaction with a multi-component field that has
a component acquire non-vanishing vacuum expectation value (plus the \emph{scalar} Higgs boson)
\cite{Weinberg,Peskin}.  The neutral components of the gauge fields furthermore mix to generate a neutral massive
vector boson ($Z$), along with the photon ($A$).
In this paper, electro-weak bosons that transform under the $\Gamma=1$ representation
of (\ref{CausalSpinorFieldEqn}) are developed. 
The matrix representations $\mathbf{\Gamma}^\beta$ are 
10-dimensional, inherently combining scalars with vectors.
For massive particles, the momentum-space forms of the $\Gamma=1$ spinors depend on
kinematic energy-momentum relationships (as in the Dirac representation), while the $m \rightarrow 0$
form of \emph{z-moving} spinors are pure numbers.  Kinematic unitary mixing of the degenerate $\gamma=0$ momentum-space
spinors will be explored, and invariant mixing angles will be developed.
Furthermore, it will be demonstrated that an appropriate co-moving spinor with a mass
consistent with that of the Higgs boson provides precisely the spinor forms needed for the mixing spinors to be
eigenstates for \emph{massive} representations of the group algebra.  
In what follows, natural units with $\hbar=1, c=1$ will be utilized, and Einstein's summation convention over repeated
\textit{super}/\textit{sub}scripts will be assumed.

\setcounter{equation}{0}
\section{Review of standard electro-weak modeling \label{Sec:EWreview}}
\indent \indent
To begin, a brief description of relevant aspects of the standard electro-weak model\cite{Peskin}
will be presented.  Consider a non-interacting field $\phi$ whose
dynamics are described using a Lagrangian density of the form
$\mathcal{L}_o =-{1\over 2}\left( \partial_\mu \phi^\dagger \eta^{\mu \nu} \partial_\nu \phi + m^2 \phi^\dagger \phi  \right)$.
This field has a real mass $m$, with momentum-space components satisfying the usual relativistic energy-momentum dispersion relation.\\
\textbf{Local gauge invariance:}
The dynamics of the field $\phi$ in $\mathcal{L}_o$  can be made invariant under unitary \emph{local} gauge transformations
$\phi(x) \rightarrow \tilde{\phi}(x)=\mathbf{U}(\alpha) \phi(x)\equiv e^{i g \,\alpha^b(x) \mathbf{G}_b}\phi(x) $ (where
the $\mathbf{G}_b$ are hermitian generators of infinitesimal transformations using group parameters $\delta\alpha^b$)
by introducing \emph{gauge potentials} $B_\mu^{(a)}$.
The gauge potentials are used to construct gauge-covariant derivatives which replace the standard gradients
$\partial_\mu \rightarrow \mathbf{D}_\mu \equiv \mathbf{1}\partial_\mu -i g\, B_\mu^{(a)}\mathbf{G}_a$,
and the potentials transform according to $B_\mu^{(a)} \mathbf{G}_a \rightarrow  \tilde{B}_\mu^{(a)} \mathbf{G}_a=
 B_\mu^{(a)}\mathbf{U}(\alpha)
\mathbf{G}_a \mathbf{U}^\dagger (\alpha) +  \partial_\mu \alpha^a \, \mathbf{G}_a$.
In addition, a term of the form ${1 \over 16 \pi}F_b^{\mu \nu} F^b_{\mu \nu}$ (using \emph{Gaussian} units) is included in the Lagrangian to generate the
field equations fully describing how charges create and interact with the gauge potentials.\\
\textbf{Spontaneous symmetry breaking}: Suppose that an interaction potential $\mathcal{V}(\phi^\dagger \phi)$ is subtracted from the
Lagrangian form $\mathcal{L}_o$.  Such a potential maintains the symmetries of $\mathcal{L}_o$, and any linear term
in the potential simply adjusts its mass term. Consider a potential of the form  $\mathcal{V}(\phi^\dagger \phi) \cong
{\lambda \over 2} (\phi^\dagger \phi)^2$ in the Lagrangian (with $\lambda>0$), i.e.
$
\mathcal{L}=-{1\over 2}\left\{ \partial_\mu \phi^\dagger \eta^{\mu \nu} \partial_\nu \phi + m^2 \phi^\dagger \phi +
\lambda  (\phi^\dagger \phi)^2  \right \}.
$
The effective potential $\mathcal{V}_{eff}\equiv {1 \over 2} \left[ m^2 \phi^\dagger \phi +
\lambda  (\phi^\dagger \phi)^2 \right]$ has a minimum at  $<\phi> \equiv \left < vac | \hat{\phi} |vac \right>=0$, indicating the
vacuum as the lowest energy (ground) state of the system.  However, if the mass becomes `tachyonic' $m\rightarrow i \mu$ , then
 $\mathcal{V}_{eff}\rightarrow {1 \over 2} \left[ -\mu^2 \phi^\dagger \phi +
\lambda  (\phi^\dagger \phi)^2 \right]$, which has a minimum when $<\phi^\dagger \phi> \cong {\mu^2 \over 2 \lambda}$, defining
a new \emph{set} of ground states for the complex field $\phi$. If the system settles into one of these ground states, the system itself
loses a symmetry of the overall Lagrangian, and gains a massless (Goldstone) mode for excitations along the un-chosen
set of ground states.  Thus, the system then manifests a \emph{broken symmetry}.

\subsubsection*{The Glashow-Salam-Weinberg model}

Weak interactions maximally discriminate between left-handed and right-handed particles.  To model this
phenomenology, the field $\phi_H$ was chosen to initially manifest local gauge symmetry under the
group of transformations $SU(2)_w \times U(1)_Y$,
where the three $SU(2)$ (weak-isospin) generators will be denoted
by $\hat{\tau_a}$, and the $U(1)$  (weak-hypercharge) generator will be denoted by $\hat{Y}$.
The Lagrangian density that manifests this local invariance takes the form
\be
\mathcal{L}_{EW}=\left( (\partial_\mu -i g_W W_\mu^{(a)} \hat{\tau}_a -i g_B B_\mu \hat{Y}) \phi \right )^\dagger \eta^{\mu \nu}
( \partial_\nu -i g_W W_\nu^{(b)} \hat{\tau}_b -i g_B B_\nu \hat{Y}  ) \phi - V(\phi^\dagger \phi)
\label{EWLagrangian}
\ee
where $V( {\phi}^\dagger \phi) \cong -\mu_\phi^2 \, \phi^\dagger \phi + \lambda (\phi^\dagger \phi)^2  $. This potential form has a minimum when
$| \phi^\dagger \phi | \cong {\mu_\phi^2 \over 2 \lambda}$ rather than $| \phi^\dagger \phi |=0$, allowing a component of the
field $\phi$ to settle into a particular non-vanishing minimal value thereby breaking its symmetry
from that of the Lagrangian.

The field is then expressed in the form of a weak iso-doublet $\hat{\phi}\equiv \left (
\begin{array}{c}
\hat{\phi}^1 + i \hat{\phi}^2 \\ 
\hat{\phi}^0 + i \hat{\phi}^3
\end{array} \right )$, where the \emph{scalar} component acquires a non-vanishing vacuum expectation value
$\left < vac | \hat{\phi}^0 |vac \right>\equiv <\phi^0>={\mu_\phi \over \sqrt{2 \lambda} }\equiv {v \over \sqrt{2}}$,
spontaneously breaking its symmetry and
defining a new ground state.  Expanding about this new ground state
$\hat{\phi}^0\equiv  {v \over \sqrt{2}}+\hat{h}$ defines a Higgs scalar field $\hat{h}(x)$ that has vanishing
vacuum expectation value and \emph{real} (time-like) mass $m_H=\sqrt{2}\, \mu_\phi$.
The remaining fields $\phi^1$, $\phi^2$, and $\phi^3$ continue to transform together under SU(2)
maintaining gauge invariance, and thus can be eliminated by a suitable choice of gauge.  These three degrees
of freedom are `consumed' in the generation of longitudinal components of what become
\emph{massive} $W^\pm$ and $Z$ bosons.

The presence of a symmetry-breaking vacuum expectation value in the Higgs sector has significant
consequences for the gauge vector potentials $W_\mu^{(a)}$ and $B_\mu$ in (\ref{EWLagrangian}). 
After substituting the form of $\hat{\phi}^0$ and examining quadratic terms for $W_\mu^{(\pm )}$ in (\ref{EWLagrangian}),
one discovers a mass term $m_{W^{(\pm)}}={1 \over 2} v |g_W|$ for this previously massless gauge boson.
Thus the weak force mediated by these bosons becomes short ranged.  The remaining gauge bosons
 $W_\mu^{(3)}$ and $B_\mu$ are then mixed to generate a massless photon that appropriately
couples to electric charge, along with a massive charge-neutral $Z$ boson.

If one substitutes the form
\be
W_\mu^{(3)}=Z_\mu \cos \vartheta_{WZ} - A_\mu^{(3)} \sin \vartheta_{WZ}, \quad
B_\mu= Z_\mu \sin \vartheta_{WZ} + A_\mu^{(3)} \cos \vartheta_{WZ},
\label{WBtoZAstandard}
\ee
into (\ref{EWLagrangian}), the proper coupling of electric charge $e$ to a massless photon field $A_\mu$
requires that $e=g_W\sin\vartheta_{WZ}=g_B \cos \vartheta_{WZ}$.  Furthermore, the term quadratic in $Z_\mu$
defines its mass as $m_Z={1 \over 2}v \sqrt{g_W^2+g_B^2}$.  Combining these equations, one arrives at the
standard electro-weak identifications:
\be
e=g_W\sin\vartheta_{WZ}=g_B \cos \vartheta_{WZ}, \quad
\cos \vartheta_{WZ}={m_W \over m_Z}, \quad
 v= {2 m_W \sin \vartheta_{WZ} \over |e|}.
\ee

\setcounter{equation}{0}
\section{Unitary mixing of degenerate momentum-space bosonic spinors}
\indent \indent
Momentum-space spinor fields described in (\ref{CausalSpinorFieldEqn})
satisfy
\be
\mathbf{\Gamma}^\beta p_\beta \,
 \mathbf{\Phi}_{\gamma,J_z}^{(1,J)} ( \vec{p})
 = -\gamma \, m \:  \mathbf{\Phi}_{\gamma,J_z}^{(1,J)} ( \vec{p}),
\label{MomentumEigenSpinorEqn}
\ee
where the spin $J$ are integers satisfying $0 \le J \le \Gamma=1$, and
$-J \le \gamma \le +J$ represent an additional set of discrete quantum numbers. 
The $\mathbf{\Gamma}^\beta$ are 10$\times$10 matrices developed elsewhere\cite{JLFQG,JLLSF13,JLNPUMBCS}.
The 10-spinors will be
\emph{Hermitian} normalized and labeled such that for a state at rest, all elements except one vanish, with ordering
$\mathbf{\Phi}^{(1,J)}_{\gamma,J_z} =\{\Phi^{(1,0)}_{0,0},\Phi^{(1,1)}_{1,1},\Phi^{(1,1)}_{1,0},\Phi^{(1,1)}_{1,-1},\Phi^{(1,1)}_{0,1},
\Phi^{(1,1)}_{0,0},\Phi^{(1,1)}_{0,-1},\Phi^{(1,1)}_{-1,1},\Phi^{(1,1)}_{-1,0},\Phi^{(1,1)}_{-1,-1}\}$.
It should also be noted that in the limit $m \rightarrow 0$, the forms
$\mathbf{\mathcal{A}}_{\gamma,J_z}^\beta \equiv \mathbf{\Gamma}^\beta \, \mathbf{\Phi}_{\gamma,J_z}^{(0,1)} ( \vec{p})$ are
momentum space representations of transverse vector fields that are degenerate with massive fields
having $\gamma=0$.  This does not occur for Dirac representation fields $\Gamma={1 \over 2}$.

\textbf{General kinematic mixing}:
It is of interest to examine (unitary) mixing of $\gamma=0$ degenerate spinors
satisfying (\ref{MomentumEigenSpinorEqn}), which does not occur amongst the $\gamma \neq 0$ spinors.
Utilizing identifications consistent with the notations reviewed in Section \ref{Sec:EWreview}, orthogonal
spinors $\mathbf{W}$ and $\mathbf{B}$ can be mixed as long as one is `tachyonic' (\emph{space-like}),
e.g. $m_B\rightarrow i \mu_B$, with
\be
p_B \rightarrow \mp {\mu_B \over m_W} \sqrt{m_W^2 + p_W^2} \quad , \quad
\sqrt{m_B^2 + p_B^2} \rightarrow \pm {\mu_B \over m_W} p_W \ge 0,
\label{pBEBeqns}
\ee
where the signs insure positivity of $\epsilon_B$. A kinematic mix vertex
is represented in Fig. \ref{MixVertex}.
\onefigure{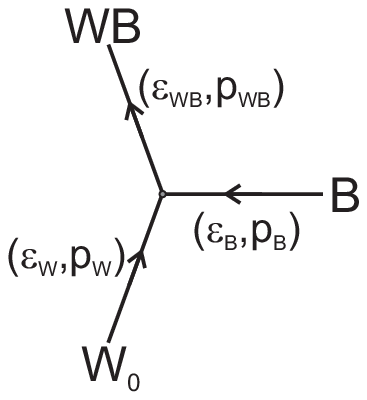}{Kinematic on-shell mixing requires $\vec{P}_W+\vec{P}_B=\vec{P}_{WB}$.}
The degenerate $\gamma=0$ spinors $\mathbf{W}_0(\vec{p}_W)$ and $\mathbf{B}(\vec{p}_B)$
can mix to generate new spinor forms $\mathbf{WB}$ and $\mathbf{BW}$ that continue
to satisfy (\ref{MomentumEigenSpinorEqn}) according to
\be
\begin{array}{r}
\cos \theta_{WB} \: \mathbf{W}_{0,J_Z} ( \vec{p}_W) +
\sin \theta_{WB} \: \mathbf{B}_{J_Z} ( \vec{p}_B)  =
\mathbf{WB}_{0,J_Z} ( \vec{p}_{WB}) , \\
-\sin\theta_{WB} \: \mathbf{W}_{0,J_Z} ( \vec{p}_W) +
\cos \theta_{WB} \: \mathbf{B}_{J_Z} ( \vec{p}_B) =
\mathbf{BW}_{0,J_Z} ( \vec{p}_{BW}),
\end{array}
\label{WBmixingEqn}
\ee
where for z-moving systems, the parameter $J_z$ labels the helicity of the particle.  The
index $\gamma=0$ on a spinor should not be confused with any co-variant space-time index.

In what follows, \emph{only} motions parallel to the z-axis will be examined,
and general momentum eigenspinors can be developed through a spatial rotation.   Labeling
the momentum of boson $X$ by $p_X$, the normalized eigen-spinors are most elegantly expressed in terms
of dimensionless \emph{kinematic angles} $\zeta_X$ defined as 
\be
\zeta_X \equiv \sin^{-1}\left (
p_X \over \sqrt{m_X^2 + 2 p_X^2}
\right ), p_X = m_X {\sin \zeta_X \over \sqrt{\cos 2 \zeta_X}},
\epsilon_X=m_X {\cos \zeta_X \over \sqrt{\cos 2 \zeta_X}}.
\label{zetamEqn}
\ee
The parameter $\zeta$ is directly related to the Lorentz boost $\beta \equiv {p \over \epsilon}=\tan \zeta$,
which implies that co-moving spinors have identical forms expressed in terms of $\zeta$.  For massive
particles, $-{\pi \over 4} < \zeta_m <+{\pi \over 4}$, while massless particles take the extremal values
$\zeta_{massless}=\pm {\pi \over 4}$, and `tachyonic' systems will satisfy ${\pi \over 4}< |\zeta_{m=i\mu}|<{\pi \over 2}$.

Degenerate spinors
moving parallel to the z-axis take a general form similar to
\be
\mathbf{\Phi}_{0,J_z=+1}^{(1,J=1)} ( \vec{p}_X) \equiv
\mathbf{X}_{0,+1} (\zeta_X)=\left (
\begin{array}{c}
0 \\ \sin \zeta_X \over \sqrt{2}  \\ 0 \\ 0 \\
\cos \zeta_X \\ 0 \\ 0 \\ \sin \zeta_X \over \sqrt{2} \\ 0 \\ 0
\end{array} \right )
\stacksymbols{\Longrightarrow}{{}_{m_X \rightarrow 0}}{4}{0.1}
\left (
\begin{array}{c}
0 \\1 \over 2  \\ 0 \\ 0 \\
1 \over \sqrt{2} \\ 0 \\ 0 \\ 1 \over2 \\ 0 \\ 0
\end{array} \right )
\ee
(with $J_z=0$ components mixing scalar and longitudinal vector elements 1,3,9,  while
$J=1,J_z=0$, has only a single non-zero component of 1 as the 6th element).
Notice that the z-moving spinors  with $m_X \rightarrow 0$ have components that are pure numbers
(though not \textit{rotationally} invariant).
Assuming only orthogonality 
$\left [\mathbf{B} (\zeta_{B}) \right ]^\dagger \mathbf{W}_0 (\zeta_{W})=0 $
and $ \theta_{WB}>0$ in (\ref{WBmixingEqn}), one can show that
if $Sign (\zeta_B)=-Sign(\zeta_W)$, then $\zeta_B=\zeta_W-Sign(\zeta_W){\pi \over 2}$ and
$Sign(\zeta_W) (\zeta_W -\zeta_{WB})=\theta_{WB}$.

The form of the parameter $\zeta_W$ can be expressed in terms of Lorentz invariants by defining
$(p_W^\mu + p_B^\mu) \, \eta_{\mu \nu} \,  (p_W^\nu + p_B^\nu) \equiv -M^2$ while using
(\ref{pBEBeqns}) to obtain
\be
\zeta W ^M _{\mu_B}= \pm \sin^{-1} \left (
\sqrt{  -2 m_W \mu_B + \sqrt{M^4+2 M^2 (\mu_B^2-m_W^2 ) +(\mu_B^2 +m_W^2)^2   }   } \over
\sqrt{2} \left ( (M^2-m_W^2)^2 + 2 (M^2 +m_W^2) \mu_B^2 + \mu_B^4    \right )^{1 \over 4}
\right ),
\label{zetaWmixEqn}
\ee
where the sign matches that of the momentum of the $\mathbf{W}_0$ spinor.
The analogous formula for $\zeta B ^M _{\mu_W}$ has a sign opposite
$\zeta W ^M _{\mu_B}$ with the labels $W$ and $B$ interchanged.  The angle $\zeta W ^M _{\mu_B}$
vanishes when $M \rightarrow \sqrt{m_W^2-\mu_B^2} < m_W$. 
The kinematic angle of the spinor $WB$ resulting
from this mixing satisfies
(assuming $p_W \ge 0$)
\be
{\zeta WB}^{M}_{\mu_B}=\sin^{-1} \left(  {p_W - {\mu_B \over m_W} \epsilon_W \over 
\sqrt{M^2 + 2 \left( p_W - {\mu_B \over m_W} \epsilon_W  \right )^2}}  \right) =\zeta W ^M _{\mu_B}-\theta_{WB} ,
\label{zetaWBeqn}
\ee
where $\epsilon_W\equiv \sqrt{m_W^2 + p_W^2}$.

\textbf{Mixing to a massless A}:
The resultant kinematic angle for mixing to a massless spinor $M\rightarrow 0$ must be $\pm {\pi \over 4}$. 
The kinematics then constrains $\tilde{m}_B \leftrightarrow i \mu_B$ such that
\be
\zeta B^0_{m_W}|_{\tilde{m}_B}  \rightarrow \tilde{\theta}_{WB} - {\pi \over 4}, \:
 \zeta W^0_{\tilde{\mu}_B}= {\pi \over 4}-  \tilde{\theta}_{WB} \Rightarrow \tilde{\mu}_B=m_W \tan \tilde{\theta}_{WB}.
\label{mBmuBeqn}
\ee
This \emph{analytically} fixes a mixing angle independent of the kinematic parameter $M$, a necessary characteristic for
quantum field mixing defined in configuration space.
With this identification, the form (\ref{zetaWBeqn}) satisfies $\zeta WB^{M}_{\tilde{\mu}_B} \goesto{M}{0} \mp {\pi \over 4}$.

\textbf{Mixing to a massive Z}:
The particular case in which the invariant mass of the WB mixture
$M=m_Z \equiv {m_W \over \cos  \tilde{\theta}_{WB}}$ \emph{analytically} generates a unique kinematic solution for which
\be
\zeta W^{m_W \over \cos \tilde{\theta}_{WB}}_{m_W \tan \tilde{\theta}_{WB}}\equiv \zeta W^{m_Z}_{\tilde{\mu}_B}
 ={\tilde{\theta}_{WB} \over 2} = -\zeta{WB}^{m_Z}_{\tilde{\mu}_B}.
\ee
Furthermore, a massive $\mathbf{H}$ spinor with the momentum
$p_H= p_W(\zeta W^{m_Z}_{m_H})$ for space-like mixing with a $W$ to produce a $Z$
is \emph{co-moving} with the $\mathbf{W}_0$ involved in the unitary mixing (as well as anti-moving with the $\mathbf{Z}$),
$\zeta_H (p_W (\zeta W^{m_Z}_{m_H}))=\zeta W^{m_Z}_{\tilde{\mu}_B}$.
The \emph{vector} components of this $\mathbf{H}$ spinor then
contribute the momentum-space spinor (and phase-space factors) to be `consumed' by
now \emph{massive} non-mixing $\mathbf{W}_{\pm 1}$ states, as well as the $\mathbf{W}_0$ mix
with $\mathbf{B}$ to the massive $\mathbf{Z}$ state.  This co-moving concurrence \emph{only}
occurs for $M={m_W \over \cos \tilde{\theta}_{WB}}$.
Thus, the relation $\zeta_H(p_W(\zeta W^{m_Z}_{m_H}))={\tilde{\theta}_{WB}\over 2}$
yields an additional kinematic condition given by
\be
{m_W \over m_Z}  =  {2 m_H^3 \over
2 m_H^3 -2 m_H  m_W^2 +m_W \sqrt{(m_H^2 +m_W^2)^2 + 2 (m_H^2 -m_W^2) m_Z^2 + m_Z^4} },
\label{CosWZHWZHEqn}
\ee
which determines another mass ratio.  It should be noted that the mixing angle defined by
$W+H\rightarrow Z$ is significantly different from $\tilde{\theta}_{WB}$.

Substitution of mass values from the \textit{Particle Data Group}
(PDG)\cite{PDG} results in a calculation of the mass of the Higgs boson well
within the experimental uncertainty, ${\Delta m_H \over m_H}\sim6.4 \times 10^{-5}$
vs. $1.1 \times 10^{-3}$.
Alternatively, the calculated mixing angle 
deviates from the quoted values by 
${\Delta \cos \tilde{\theta}_{WB} \over m_W/m_Z }\simeq 6.3 \times 10^{-6}$, also
well within the experimental uncertainty of $1.5 \times 10^{-4}$.

\setcounter{equation}{0}
\section{Additional kinematic correspondences}
\indent \indent

As asserted in the previous section, the Higgs scalar is the only component of the
$\mathbf{H}$-spinor not consumed to produce massive vector bosons.  These components
have momenta satisfying $p_H (\zeta W^{m_Z}_{\tilde{\mu}_B})=p_W(\zeta W^{m_Z}_{m_H})$. 
It is of interest to next explore analogous relationships involving \emph{energies}. 

At a fundamental level, one expects the $\mathbf{Z}$ spinor to have the potential to
directly mix into other degenerate states.  In particular, examine the
mixing of a $\mathbf{Z}$ with an orthogonal $\mathbf{\phi}$ (whose symmetry was spontaneously broken)
having `tachyonic' mass $\mu_\phi={m_H \over \sqrt{2}}$.  The $\mathbf{Z}$ should be able
to \emph{self}-generate $Z+\phi\rightarrow Z$, with kinematic angle given by $\zeta Z ^{m_Z}_{\mu_\phi}$,
calculated from (\ref{zetaWmixEqn}). Furthermore,
using (\ref{mBmuBeqn}) this can define a particular $Z+\phi$ mixing angle into a massless state as
\be
\tilde{\theta}_{Z \phi}\equiv \tan^{-1}{\mu_\phi \over m_Z}=\tan^{-1}{\mu_H \over \sqrt{2} \, m_Z}
\Rightarrow \zeta Z^M _{m_Z \tan \tilde{\theta}_{Z \phi}}=  \zeta Z ^{M}_{\mu_\phi}.
\label{ZphiMixAngle}
\ee
More generally, $\theta_{Z \phi}$ represents the Lorentz invariant mixing angle of
$\mathbf{Z}+\mathbf{\Phi} \rightarrow \mathbf{X}_M$ for all invariant energies $M$. 
Relationships of kinematic angles involving known particles ($W,\: Z$, $H$) that are co-moving
with $\zeta Z ^{M}_{\mu_\phi}$ will be examined for further insights.

In particular, 
from (\ref{zetamEqn}), the kinematic angle for a mass $m$ with energy $\epsilon$ generally satisfies
\be
\zeta_m^\epsilon (\epsilon) =\sin^{-1}\sqrt{\epsilon^2 -m^2 \over 2 \epsilon^2 -m^2}.
\label{zetaEpsilonEqn}
\ee
Consider a particle with mass  $m_*$ having threshold rest-energy for particle $m_{th}$,
which has a spinor with kinematic angle
\be
\zeta_{m_*}^\epsilon (m_{th})=\sin^{-1}\sqrt{m_{th}^2 -m_*^2 \over 2 m_{th}^2 -m_*^2},
\label{zetamstarEqn}
\ee
establishing a minimal kinematic relationship between the masses.
For the general mixing $m+X\rightarrow M_{m X}$, the
value of $ M_{m X}$ in $\zeta m^{M_{mX}}_{m \tan \tilde{\theta}_{mX}}$ 
for which it co-moves with $\zeta_{m_*}^\epsilon (m_{th})$ in (\ref{zetamstarEqn})
is obtained using (\ref{zetaWmixEqn}), yielding
\be
M_{m X}(m_*,m_{th};\tilde{\theta}_{mX})={ \sqrt{m^2 m_*^2 \cos (2 \tilde{\theta}_{m X}) +
2 m_{th}^2 \sqrt{m_{th}^4-m^2 m_*^2}\, \sin (2 \tilde{\theta}_{m X})} \over m_* \cos \tilde{\theta}_{m X}},
\label{GenMstarEqn}
\ee
where $\tilde{\theta}_{m X}$ is the previously defined mixing angle of $m+X$
that also generates a massless partner.

As an example of $Z+\phi$ mixing, consider a $\mathbf{Z}$ spinor that is co-moving with
a $\mathbf{W}$ spinor at the threshold energy of a $Z$.  One can
\emph{analytically} show that the following kinematic parameters are equal:
\be
\zeta Z^{M_{Z \phi}^{W_*}}_{\mu_\phi}=\zeta_W^\epsilon (m_Z)=\zeta_Z^\epsilon ({m_Z^2 \over m_W})=
\zeta_H^\epsilon ({m_H m_Z \over m_W})
\label{ZphiWtoZeqn}
\ee
where $M_{Z \phi}^{W_*}\equiv M_{Z \phi}(m_W,m_Z;\tilde{\theta}_{Z \phi})$.
Thus, (\ref{ZphiWtoZeqn}) generates no new kinematic condition.

Next, consider the relationship involving a $W$ at $H$ threshold, $\zeta_W^\epsilon (m_H)=
\zeta_H^\epsilon ({m_H^2 \over m_W})$, which is analytically true.
For $Z+\phi$ mixing, a $Z$ satisfying the \emph{energy} analog of
$\zeta_H (p_W (\zeta W^{m_Z}_{m_H}))=\zeta W^{m_Z}_{\tilde{\mu}_B}$
in $W+B$ mixing has
\be
\zeta_H^\epsilon (\epsilon_Z ={m_H^2 \over m_W})\equiv \zeta Z^{M_*}_{\mu_\phi}\quad \Rightarrow \quad
 \zeta Z^{M_*}_{\mu_\phi} = \zeta_W^\epsilon (m_H),
\ee
providing an invariant $M_*$ that can be connected to the previous $W+B$ kinematic mixing.  Using symmetry, the kinematic angle
for the $W+B$ mixing satisfies
\be
\zeta_Z^\epsilon ({m_H^2 \over m_W})=\zeta_W^\epsilon ({m_H^2 \over m_Z})\equiv 
\zeta W^{M_*}_{m_W \tan \tilde{\theta}_{WB}},
\ee
which determines $M_*$.  Reversing the previous step using the analytic relationship
$\zeta_H^\epsilon (\epsilon_W={m_H^2 \over m_Z})=\zeta_Z^\epsilon(m_H) $
defines the invariant $M_*$ as
\be
M_*=M_{WB}(m_Z,m_H;\tilde{\theta}_{WB}) \sim 245.7 GeV .
\label{MstarValue}
\ee
However, unlike the first equality in (\ref{ZphiWtoZeqn}),
the relationship $\zeta Z^{M_*}_{m_\phi} \approx \zeta_W^\epsilon (m_H)$ is \emph{not} an analytic identity,
differing in the $6^{th}$ significant digit from PDG\cite{PDG} values, thus establishing an additional kinematic condition.

Numerically solving this new relationship
\be
\zeta Z^{M_*}_{m_\phi}=\zeta_W^\epsilon (m_H)
\ee
using (\ref{zetamstarEqn}), (\ref{GenMstarEqn}), and (\ref{MstarValue}), along with
$\cos \tilde{\theta}_{WB}={m_W \over m_Z}$ and (\ref{CosWZHWZHEqn}),
determines all mass ratios, leaving only
a single undetermined mass scale.
The resulting mass ratios
\be
\begin{array}{l}
\tilde{\theta}_{WB}=0.49183562809461... \\
{m_W \over m_Z}= 0.88146747846957...\\
{m_H \over m_Z}=1.37196507659589... \\
\end{array}
\label{JLvaluesEqn}
\ee
all fall \emph{well} within the uncertainties in measured values:
${\Delta \cos \tilde{\theta}_{WB}\over m_W/m_Z }\simeq 1.2\times 10^{-6}$ vs. experimental uncertainty
$1.7 \times 10^{-4}$, and $m_H\simeq 125.106 \, GeV$ vs. $125.1 \pm 0.14 \, GeV$.
 It should be noted that the previously reported value\cite{JLNPUMBCS} for the mixing angle of a
(slightly different) self-generating system $\theta_{WB}$ differs
from $\tilde{\theta}_{WB}$ at the 6$^{th}$ significant figure, making them inconsistent.  It is felt that
the value presented here is more fundamental and robust.


\section{Discussion and Conclusions}
\indent \indent 
Kinematic constraints relating electro-weak boson masses in terms of a single mass scale have been demonstrated. 
These constraints follow from unitarily mixing momentum-space spinors that transform as the first boson states 
beyond the fundamental fermion (Dirac) spinors,
with the scalar component of an $\mathbf{H}$ spinor being associated with the Higgs scalar. 
The vector components of the  $\mathbf{H}$ spinor are `consumed' to generate \emph{massive} spinor forms
for the $\mathbf{W}^\pm$ and $\mathbf{Z}$.
The presented constraints fall well within the reported experimental uncertainties.

The developed relationships are consistent with the standard mixing of space-time dependent fields in Lagrangian dynamics,
which can depend only on kinematic invariants. 
However, these results do not imply that dynamic particle couplings are independent of four-momentum transfers. 
Both non-perturbative unitary (vertex) structure functions, as well as renormalized perturbative effective couplings
are typically \emph{not} kinematically invariant.
It should be emphasized that this formulation does not contradict the Standard Model, but rather
provides supplementary insight into the Higgs sector of that model.

Furthermore, any bosons of mass $M$ resulting from $\zeta Z^M_{\mu_\phi}$ mixing do \emph{not}
couple to standard electromagnetic charge.  Although the kinematic invariant energy $M_* \sim 245.7 GeV$
was developed as a convenience for connecting $Z+\phi$ mixing to $W+B$ mixing
through this shared value, any \emph{physical} massive or massless particle resulting from $Z+\phi$
mixing would provide suggestive and intriguing
inferences of electromagnetically-dark sector dynamics.
Promising expressions relating particle-gauge field couplings to weak boson kinematics
are actively being examined as on-going research.


\end{document}